\documentstyle[aps,psfig]{revtex}

\def\be{\begin{equation}}
\def\ee{\end{equation}}
\def\bea{\begin{eqnarray}}
\def\eea{\end{eqnarray}}

\begin{document}
%%%%%%%%%%%%%%%%%%%%%%%%%%%%%%%%%%%%%%%%%%%%%%%%%%%%%%%%%%%%%%%%%%%%
\draft

\title{Coherent Line Removal:\\
Filtering out harmonically related line interference from experimental 
data, with application to gravitational wave detectors}

\author{Alicia M. Sintes and  Bernard F. Schutz}
\address{Max-Planck-Institut f\"ur Gravitationsphysik
(Albert-Einstein-Institut) \\
Schlaatzweg 1, 14473 Potsdam, Germany}
\date{\today}
 \maketitle

\begin{abstract} 
We describe a new technique  for removing troublesome interference
from external coherent signals present in the gravitational wave spectrum.
The method works when the interference is present in many harmonics,
as long as they remain coherent with one another.
The method can remove interference even when the frequency changes.
We apply the method to the data 
produced by the Glasgow laser interferometer in 1996 and the entire series of
wide lines corresponding   to the electricity supply
frequency and its harmonics
 are removed, leaving the spectrum clean enough to
 detect  possible signals previously masked by them.
We also study the effects of the line removal 
on the statistics of the noise in the time domain.
We find that this technique seems to reduce  the level of
 non-Gaussian noise present in the interferometer and therefore,
 it can raise the sensitivity and duty cycle of the detectors.
\end{abstract} 
 \pacs{04.80.Nn, 07.05.Kf, 07.50.Hp, 07.60.Ly}

\section{Introduction} 
In this paper we present a new  procedure to remove external
interference from the output of 
gravitational wave detectors. 
This method allows the removal  of phase-coherent lines, not stochastic ones
(such as those due to thermal noise), while keeping the intrinsic 
detector noise.
The method works so well that real gravitational wave signals masked
by  the interference can be recovered.

Our method requires 
coherence between the fundamental and several harmonics. If there is no
such coherence, other methods are available \cite{A,Th,p3}, 
but these methods will remove real signals as well. 
It is \lq safe' to apply this technique to gravitational wave data
because we expect that coherent gravitational wave signals 
will appear with at most the fundamental and one harmonic
\cite{bo}.
The information in the harmonics of the interference  can be used to 
removed it without disturbing \lq single-line' signals.
Therefore, this method can
be very useful in the search for monochromatic {\sc gw} 
signals, such as those  produced by pulsars
 \cite{S,T}.

The method can be used to remove 
periodic or broad-band
signals (e.g.,  those which change frequency in time), 
provided their harmonics are sufficient strong and numerous, even if
there is no external reference source. The method requires little or
no a priori  knowledge of the signals we want to remove. This is a
characteristic that distinguishes it from other methods such as
adaptive noise cancelling ({\sc anc}) \cite{anc}, which makes use of an
auxiliary reference input derived from one or more sensors. 
Although in some cases {\sc anc} can be used without a reference input, it is
not clear how  {\sc anc} can, in those cases, cancel harmonics of
broad-band signals and at the same time detect weak periodic signals
masked by them. 
This is particularly important in gravitational wave detection,
where signals can be as high as 2 kHz while the interference has a
fundamental frequency of 50 or 60 Hz.

In this paper, we illustrate the usefulness of this new  technique 
by applying
it to the data produced by the Glasgow laser interferometer in March 1996
and removing all those lines  corresponding to the electricity supply
frequency and its harmonics.
As a result the interference is attenuated or eliminated by cancellation
in the time domain and the power spectrum appears completely clean allowing
the detection of signals that were buried in the interference.

The removal improves the data in the time-domain as well. 
Strong interference produces a significant  non-Gaussian component
to the noise. Removing it therefore improves the
sensitivity  of the detector to short bursts of {\sc gw}'s. 

The rest of the paper is organized as follows. In the next section
we define {\it coherent  line removal} and we give an algorithm for it.
In section III we apply  the method to remove the 50 Hz harmonics from
{\sc gw} interferometric data.
First, we focus our attention  to the performance of the algorithm on
two minutes of data and we check what happens to a signal masked by
the interference. Then, we implement an automatic procedure to clean
the whole data stream. We have to solve some difficulties such as 
the presence of small gaps in the data.
We present the results obtained and we show how the electrical interference
is completely removed.
Moreover, we study the  effects of the line removal  on the
statistics of the noise in the time domain.
We compare  the mean, the standard deviation, the skewness and the kurtosis
of the data before and after the line removal an we
 also study the Gaussian character. We  apply 
 two statistical test to the data: the chi-square test and the 
 Kolmogorov-Smirnov test.
 Finally, in section IV we  discuss the results obtained.

\section{The principle of  Coherent  Line Removal}

In this section, we define {\it coherent line removal} and  give
an algorithm for it.
  
  We suppose that a certain interference signal $m(t)$ 
  (e.g., 50 Hz interference
from the main electricity supply) enters  the system. 
It  may already contain harmonics, and non-linear effects in the system 
electronics may introduce further harmonics. If the processes that 
produce  the harmonics are stationary, then we expect the phase of the
harmonics  to be simply related to that of $m(t)$. In particular, we 
assume that the interference has the form
\be
y(t)=\sum_n a_n m(t)^n + \left( a_n m(t)^n\right)^* \ ,
\label{e3}
\ee
where $m(t)$ is a nearly monochromatic function, $n$ are natural numbers,
 and $a_n$ are
complex constants that 
depend on the processes that generate the harmonics, and which are 
not known a-priori. We suppose that $m(t)$ is a narrow-band function near
a frequency $f_0$. Then, the $n$th harmonic will be near $nf_0$.

The total output of the system  also contains noise and possibly other signals. 
Usually, the data recorded is band-limited, since an 
anti-aliasing filter is applied to the data  before it is 
sampled. Therefore, the function $y(t)$ must be band-limited as well,
and  the number of harmonics  to be considered is finite. 
This number, $n_{max}$, is given  by the Nyquist frequency and the 
frequency of the fundamental harmonic $f_0$
\be
n_{max}=f_{Nyquist}/ f_0 -1 \ .
\ee

The key to this method is to estimate the interference by using many
harmonics of the interference signal and to construct a function
\be
h(t)=\sum_{n=1}^{n_{max}} \rho_n M(t)^n + \left( \rho_n M(t)^n\right)^* \ ,
\label{e5}
\ee
that is as close a replica as possible of $y(t)$. This function is then
subtracted from the output of the system   cancelling 
the interference. 
If there is a narrow gravitational wave signal within a frequency band 
obscured by a particular harmonic, it can still be present after line
removal, because it will not match the form of the signal being removed.
Moreover, because $h(t)$ is constructed from many frequency bands with 
independent noise, the statistics of noise in any one band are little
affected by coherent line removal.

Therefore, we have to design  an algorithm  
to determine
the complex function $M(t)$ and all the parameters $\rho_n$ that
minimize the total output power. Notice that, from the experimental data, 
we do   not independently know the value of the input signal $m(t)$.

As pointed out before, we assume that the data produced by the system
is just the sum of the interference plus  noise
(we ignore here any signals not of the form (\ref{e3}))
\be
x(t)=y(t)+n(t) \ ,
\label{e6}
\ee
where $y(t)$ is given by Eq.~(\ref{e3}) and the noise $n(t)$ in the
detector  is a zero-mean stationary
stochastic process. The Fourier transform
of the data $\tilde x(\nu)$ is simply given by
\be
\tilde x(\nu)=\tilde y(\nu)+\tilde n(\nu) \ .
\label{e7}
\ee

Choosing  a subset of harmonics, $\{k\}$,
or all of them if one prefers, the idea is to construct the function
$M(t)$ by extracting  the maximum information from the harmonics considered.
The procedure consists in determining the upper and lower frequency 
limit of each harmonic considered,
 $(\nu_{ik}, \nu_{fk})$,  and  defining a set of functions
$\tilde z_k(\nu)$ in the frequency domain as
\be
\tilde z_k(\nu)\equiv \left\{
\begin{array}{cc}
\tilde x(\nu) & \nu_{ik}<\nu <\nu_{fk}\\
0 & \mbox{elsewhere}\ .
\end{array}
\right.
\label{e8}
\ee
Comparing Eq.~(\ref{e8}) with (\ref{e3}) and (\ref{e7}) we have
\be
\tilde z_k(\nu)= a_k \widetilde{m^k}(\nu) +\tilde n_k(\nu) \ ,
\ee
where
\be
\tilde n_k(\nu)= \left\{
\begin{array}{cc}
\tilde n(\nu) & \nu_{ik}<\nu <\nu_{fk}\\
0 & \mbox{elsewhere}\ ,
\end{array}
\right.
\ee
is  a zero-mean stationary  random complex noise.
Then, we calculate  their inverse Fourier transform
\be
z_k(t)=a_k m(t)^k +n_k(t) \ .
\ee
Since we assume $m(t)$ to be a narrow-band function near a frequency $f_0$,
each $z_k(t)$ is a  narrow-band function near $kf_0$. It corresponds to 
one of the harmonics of the external interference plus noise
which is independent  in the different frequency bands considered
($n_k(t)$ for different $k$), since it is 
 produced by an stationary process.
 
Now, we define\footnote{In order to perform this operation numerically, we
separate $z_k(t)$  into amplitude and  phase 
($z_k(t)=A_k(t)\, \exp(i \Phi_k(t))$, where $A_k(t)$ and $\Phi_k(t)$ are
real functions).
 Then, we correct the phase
 angle by adding multiples of $\pm 2\pi$ in order to make it continuous
 and prevent branch cut crossing. And, we construct $B_k(t)$ as
 $B_k(t)=A_k(t)^{1/k}\, \exp(i \Phi_k(t)/k)$.}
\be
B_k(t)\equiv \left[ z_k(t)\right]^{1/k}\ ,\label{e10a}
\ee 
that can be rewritten as 
\be
B_k(t)= (a_k)^{1/k}m(t) \beta_k(t) \ ,
\ee
where
\be
\beta_k(t)=\left[ 1+ {n_k(t) \over a_k m(t)^k}\right]^{1/k} \ .
\label{e10}
\ee
All these stochastic functions $\{B_k(t)\}$ are almost monochromatic around the 
fundamental frequency, $f_0$, but they have different  mean values
\be
\langle B_k(t)\rangle= (a_k)^{1/k}m(t) \ .
\ee
They differ  by a certain complex amplitude.
 In order to  estimate the interference $M(t)$,
 we need to define another set of functions
 \be
 b_k(t)\equiv\alpha m(t) \beta_k(t) \ ,
 \ee
such that, these
new functions $\{b_k(t)\}$ form a set of random variables $-$functions 
of
time$-$ and  they all have the same mean value
\be
\langle b_i(t)\rangle=\alpha m(t) \ .
\ee
The functions $b_k(t)$ are constructed  multiplying the previous functions
$ B_k(t)$ by a certain complex amplitude $\Gamma_k$
\be
 b_k(t)=\Gamma_k B_k(t)\ . 
 \ee
 The values of $\Gamma_k$ can be  obtained 
 from a least square method,  comparing
 the first harmonic considered, $k(1)$, with the other ones
 (i.e., $\Gamma_k$ are the values that minimize 
 $\vert  B_{k(1)}(t)-\Gamma_k B_k(t)\vert^2$ for each $k$).
We find
\be
\Gamma_k=
 \sum_j B_{k(1)}(j\Delta t) B_k(j\Delta t)^*{\Big
 { /}}
 \sum_j \vert B_k(j\Delta t)\vert^2 \ ,
\ee
where $\Delta t$ corresponds to the time interval between two samples,
so that the sampling frequency is equal to $1/\Delta t$, $*$ denotes the
complex conjugate, and the index  $j$ counts  
sampled points in the time domain.

Now, we want to construct $M(t)$ as a function of all $\{b_k(t)\}$,
 in such a way
that it has the same mean and minimum variance. If we assume the function
$M(t)$ to be linear with $\{b_k(t)\}$,  statistically the best estimate is
\be
 M(t)=\left(\sum_k {b_k(t) \over {\rm Var}[\beta_k(t)]} \right) {\Big
 { /}}
\left( \sum_k {1 \over {\rm Var}[\beta_k(t)]}\right) \ .
\ee
The variance of $\beta_k(t)$ can be estimated by doing a Taylor 
expansion of equation (\ref{e10}), hence we obtain
\be
{\rm Var}[\beta_k(t)]= {\langle n_k(t) n_k(t)^*\rangle\over  k^2
\vert a_k m(t)^k\vert^2}+ \mbox{corrections} \ ,
\label{eq19}
\ee
where we can  approximate
\be
\vert a_k m(t)^k\vert^2 \approx \vert z_k(t)\vert^2 \ ,
\ee
and the numerator can be rewritten as
\be
\langle n_k(t) n_k(t)^*\rangle=\int d\nu \int d\nu'
\langle \tilde n_k(\nu) \tilde n_k(\nu')^*\rangle e^{2 \pi i(\nu-\nu')t} \ .
\ee
In the case of stationary noise (i.e., 
$\langle \tilde n(\nu) \tilde n(\nu')^*\rangle= S(\nu) \delta(\nu-\nu')$),
the previous equation becomes
\bea
\langle n_k(t) n_k(t)^*\rangle & = &
\int d\nu \int d\nu' \, S_k(\nu) \delta(\nu-\nu') e^{2 \pi i(\nu-\nu')t}
\nonumber\\
& =& \int_{\nu_{ik}}^{\nu_{fk}} S(\nu) d\nu \ ,
\eea
where $S(\nu)$ is the power spectral density of the noise.

After we have estimated the interference $M(t)$, 
it only remains to determine the parameters $\rho_n$ in Eq.~(\ref{e5}).
They can be obtained
 by minimizing the total output power in the time domain, 
 $\vert x(t)-h(t)\vert^2$ (i.e., applying a
 least square method). Taking into
 account that the noise near the different harmonics is uncorrelated, it
yields
\be
\rho_n=
 \sum_j x(j\Delta t) M^n(j\Delta t)^*{\Big
 { /}}
 \sum_j \vert M^n(j\Delta t)\vert^2 \ ,
 \label{e21}
\ee
which  is the form we use in our algorithm. One can also
minimize the total output power in the frequency domain,
 $\vert \tilde x(\nu)- \tilde h(\nu)\vert^2$. Then, we get
\be
\rho_n=
 \sum_j \tilde x(j\Delta\nu) \widetilde{M^n}(j\Delta\nu)^*{\Big
 { /}}
 \sum_j \vert \widetilde{M^n}(j\Delta\nu)\vert^2 \ ,
\ee
where $j$ is now the frequency index.
%%%%%%%%%%%%%%%%%%%%%%%%%%%%%%%%%%%%%%%%%%%%%%%%%%%%%%%%%%%%%%%%%%%%%%%
\section{Removal of 50 Hz harmonics from interferometric data stream}

In this section, we present experimental results that demonstrate
the performance of the coherent line removal and
 show its potential value. This new technique  is applied
to the data produced by the Glasgow laser interferometer and the
electrical interference is successfully removed.

In the study of the data produced 
 by the Glasgow laser interferometer in March 1996
\cite{J},
we observe in the spectrum many instrumental lines, some of them at
multiples of 50 Hz. All these lines are rather broad,
 and when we compare them,
we observe that their overall structure  is very similar 
but only  the scaling of the width is different (see figure 1).

\begin{figure}[h!]
\centerline{\vbox{ 
\psfig{figure=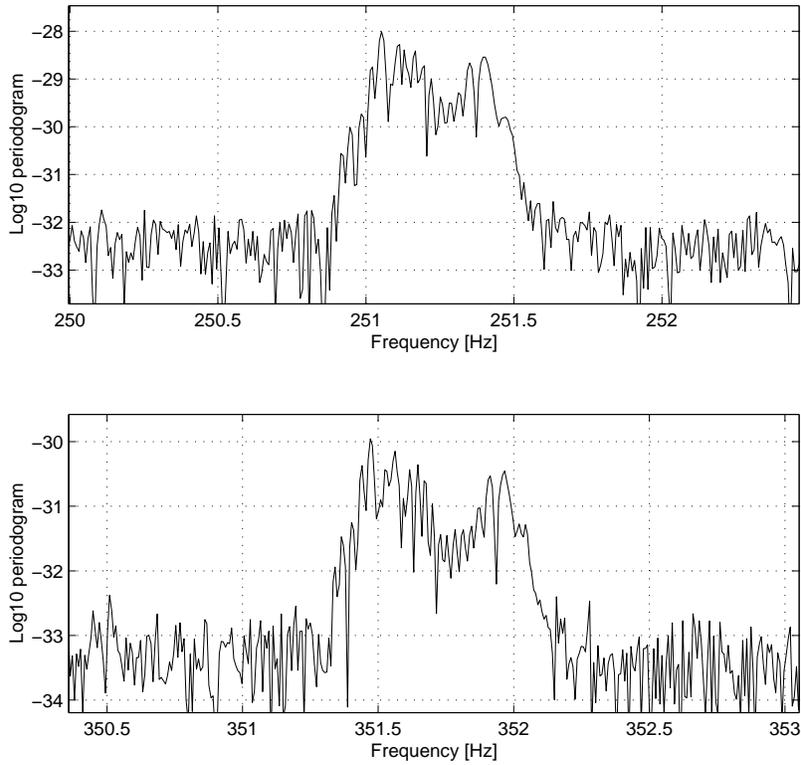,width=4.2in} 
}} 
\vspace*{10pt}
\caption[]{Comparison of the structure of the lines at 250 Hz and at 350 Hz
of the power
spectrum (of  2 minutes) of the Glasgow data.  
The broad shape is due to the wandering of the incoming electricity 
frequency.
} 
\vspace*{10pt}
\end{figure} 
If we look at these lines in more detail, in smaller length Fourier transform
(seconds in length), they appear as well defined
small bandwidth lines which change frequency over time
in the same way, while other ones
 remain at constant frequency. Therefore, all these lines at multiples
of 50 Hz must be harmonics of a single source (for example the electricity
 supply),
and this makes  them   good candidates to test our algorithm.

The same  phenomenom is observed in the Garching 30-meter prototype
\cite{ga}. 
The LIGO group has also reported largely instrumental artifacts
 at multiples of the 60 Hz line frequency in their  
 40-meter interferometer \cite{Abra}. Therefore, this seems to be a general
  effect
 present in the different prototype interferometers.
 
 In the Glasgow data, the lines at 1 kHz have a width of
5 Hz. Therefore, we can ignore these sections of the power spectrum or
we can try to remove this interference in order to be able to detect
{\sc gw} signals masked by them.

In the large-scale detectors now under construction,
we expect the amount of interference to be smaller. 
Prototypes like the Glasgow instrument were
designed to test optical systems, not to collect
 high
quality data, and the effort required to exclude interference
was not justified by the goals of the prototype development.
 However, we cannot be sure that such interference will be completely absent
or that other
sources of interference will not manifest themselves in long-duration
spectra. Indeed, the Glasgow data \cite{J} contain other regular
features of unknown origin \cite{K}. 
For this reason we 
 investigate a solution to this problem  using the existing data.

\subsection{Testing procedure}

As a first test, we  apply the 
coherent line removal  to a set of $2^{19}$ points
(approximately two minutes) of the Glasgow data. 
The stretch of data selected corresponds to a period of time in which the
detector is on lock and the level of noise is low.

After calculating the DFT of this piece of data (using the FFT algorithm),
we notice that the odd harmonics of the 50 Hz line are much stronger than
the even ones. Thus, in order to construct  the reference wave-form $M(t)$, 
we choose a set of ten harmonics $k=\{3,5,7,\ldots ,21\}$, corresponding to 
the lines 
at frequencies $50\times k(i)$ Hz. We also give as inputs the corresponding
upper and lower  frequency limits  of each of these harmonics, 
$(\nu_{ik},\nu_{fk})$,
 that are obtained from
the power spectrum. 

The first difficulty that arises is to estimate the power spectral density
({\sc psd}) of the noise (without the external electrical interference)  
in those frequency
bands corresponding  to the harmonics considered. In the absence of
any extra information, we  assume  the {\sc psd} is constant
in each frequency band  and we calculate its value    by 
averaging  the {\sc psd} in the nearness  of the line considered.
With all this information,
we apply  the line removal and we successfully remove
 the electrical interference as it can be seen in figure 2.

\begin{figure} 
\centerline{\vbox{ 
\psfig{figure=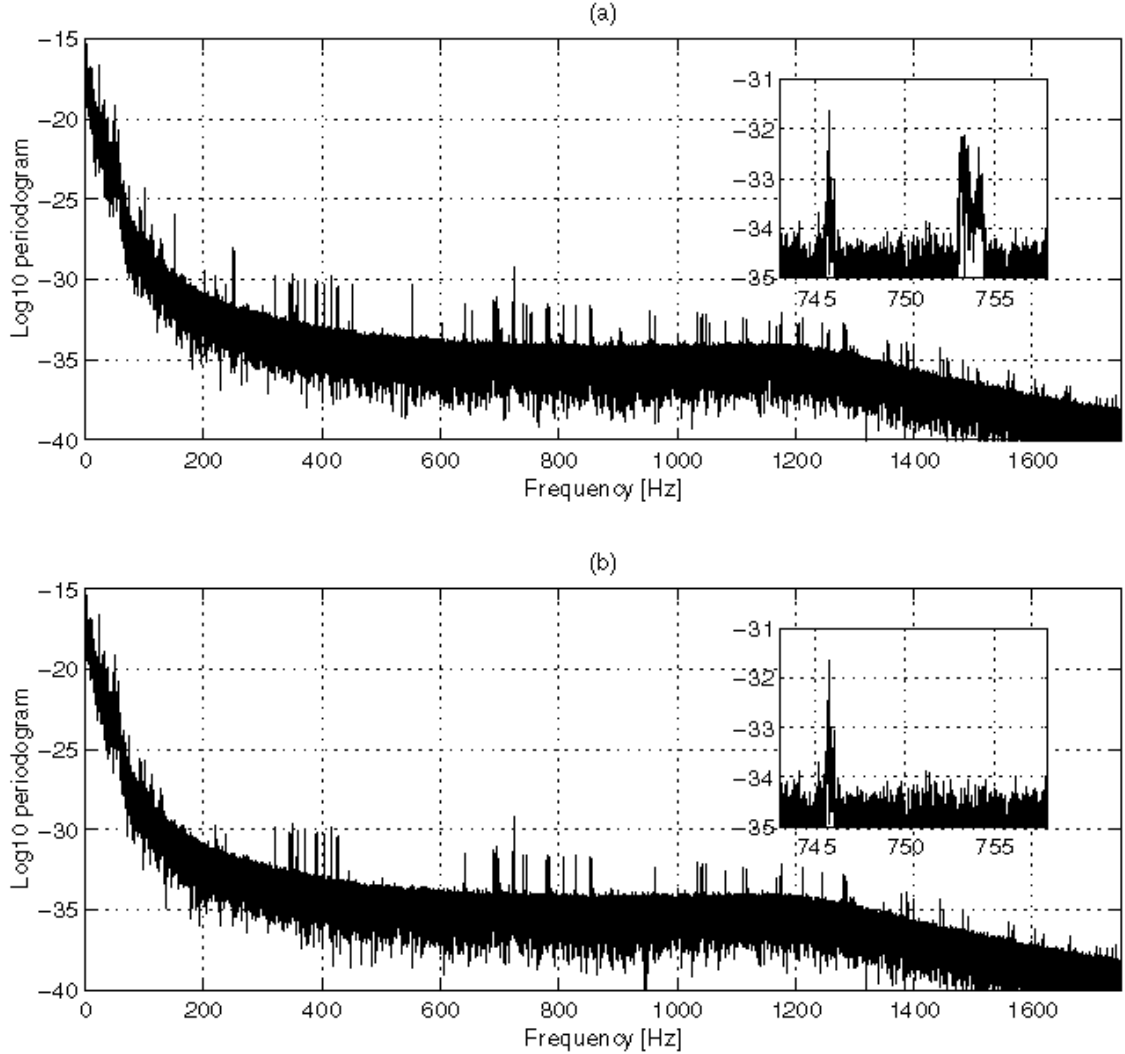,width=6.0in}
}} 
\vspace*{10pt}
\caption[]{Decimal logarithm of the
periodogram of $2^{19}$ points (approximately 2 minutes) of 
(a) the Glasgow data with the detail of 
one of the harmonics near 754 Hz.
(b) The same data after the removal of the electrical 
interference as described 
in the text. 
} 
\vspace*{10pt}
\end{figure} 

We have repeated the procedure considering only  six 
harmonics, $k=\{3, 5, 7, 9, 11, 13\}$ and, surprisingly, 
the interference  still remains below the noise level
showing  the power of this algorithm. 
The method requires further study (i.e., with simulated noise and
different signals) in order to characterize how its performance
depends upon the number of harmonics and their strength.
Of course, in order to obtain a minimum variance for $M(t)$, the
larger number of harmonics considered the better. The minimum number
of harmonics required depends, in each case, on the strength of the
harmonics considered, more precisely, with the ratios
$ \langle n_k(t) n_k(t)^*\rangle /
\vert a_k m(t)^k\vert^2$, appearing in equation (\ref{eq19}), 
which should be smaller
than one.

Now, we are interested in studying  what will happen to a signal masked
by the interference. For this purpose, we  apply  coherent line removal
  to the true
experimental data with an external simulated signal at 452 Hz, that
is initially hidden due to its weakness.

\begin{figure} 
\centerline{\vbox{ 
\psfig{figure=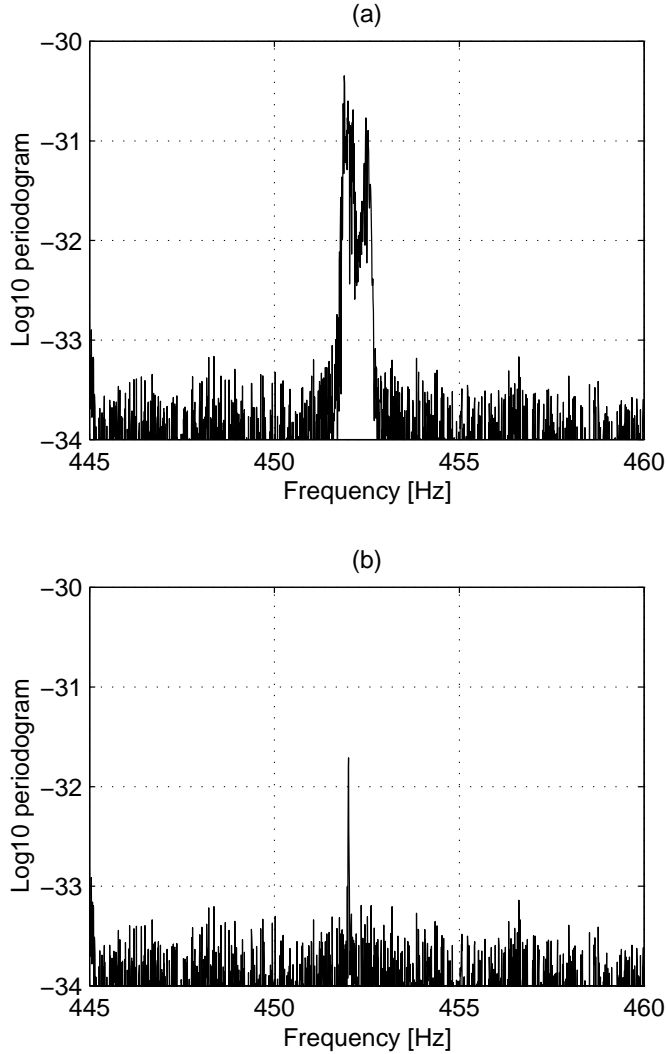,width=3.5in} 
}} 
\vspace*{10pt}
\caption[]{
(a)  Zoom of the same experimental  data as in figure 2
with an artificial  signal added at 452 Hz. 
(b) The data in (a) after  removing  the electrical 
interference, revealing that
the signal remains detectable. 
} 
\vspace*{10pt}
\end{figure} 

First, we take into account  not to consider the harmonic near 450 Hz as 
an input to estimate the interference $M(t)$. In this case, we
succeed in removing the
electrical interference while keeping the signal present in the data,
obtaining a clear outstanding peak over the noise level  and whose
amplitude is not modified
(see figure 3). 

However, now we apply the same  procedure, but including
  the harmonic near 450 Hz ($k=9$). In this case, 
$M(t)$  contains  a low-level of signal component in addition to 
the electrical interference. This  
signal component
causes
a 
cancellation 
of the $25.4\ \%$
of the amplitude of the line at 452 Hz and also the appearance
of weaker signals  under the other harmonics.
The amount of signal distortion depends basically  on the variance
of $\beta_9(t)$ with respect to the other ones.

Any of the harmonics considered 
in the construction of the reference function $M(t)$ may contain a 
\lq single-line' signal we want to recover, but a-priori,
 we will not be able 
to say  which one is masking a signal.
The key  is that gravitational wave signals will not be present with
multiples harmonics. Therefore, after applying the line removal, we
 will have a  candidate line. If
we are interested, we can repeat the procedure  without using the 
\lq\lq dangerous" harmonic and hence, we can make use of the 
whole power of the algorithm. With this aim, it is better to consider the
lower harmonics, since their frequency bands are narrower and also 
the probability of having a thermal line buried  with them 
may accordingly be lower.

\subsection{Automatic cleaning of whole data stream}
The next step consists  in removing the electrical interference  
from the whole data stream
making use of an automatic procedure.

The Glasgow data consist of 19857408 points, sampled at 4 kHz and 
quantized with 12 bit analog-to-digital converter with a dynamical range
from -10 to 10 Volts. The data are divided  into 4848 blocks of 
4096 points. For each of these blocks, the standard deviation  is 
calculated and, depending whether it exceeds a certain value, the
whole block is classified as \lq\lq bad".

The first 18 minutes of data are rendered useless
due to a failure  of autolocking. Therefore, we decide to ignore the
first 1153 blocks. The rest of the data are separated into groups of
64 blocks (approximately one minute of data) and, for each of them, 
the coherent line removal algorithm is applied.

The first difficulty that arises is  how to deal  with the \lq\lq bad"
blocks. In some cases, the detector is out of lock and, in other ones,
the level of noise is very high. Since this method  is based on 
Fourier transforms,  the suggestion of defining a window function  that 
vetoes  the \lq\lq bad" data produces that  each real feature in the
spectrum be accompanied by horrible side lobes and satellites features, 
the structure of which depends in detail on the distribution of those gaps.
In order to reduce these effects, we realize that it is better
 not to remove the \lq\lq bad"
data, but to divide each \lq\lq bad" block  by its standard deviation.
 Of course,  the \lq\lq bad"
blocks are not  taken into account in order to determine the parameters
$\rho_n$ given by Eq.~(\ref{e21}). This is done   multiplying
the function $M(t)$  and the input data  by a window function, such that,
it is equal to zero  for each block of \lq\lq bad" data and one otherwise.
As a result, all \lq\lq bad" data are set to zero in the output.

Another important  issue  is to design an strategy  to detect
the harmonics of the electricity supply interference,
 and  determine  their upper and
lower frequency limits without  interacting  with the program.
This is not an easy task, since   the noise level is
very high, the lines are broad, they vary in time, and they are
 partially hidden  by the noise. 
Also, the presence of violin modes closed to 
the harmonics, or completely overlapping with them, makes  it more
difficult.

The method we have used consists  in computing the power spectrum 
for each piece of the
 data (with the electrical interference) and we
search for the location of the 
first harmonic considered. We  calculate the mean,
$m_{k(1)}$, and the standard deviation, $\sigma_{k(1)}$, in a certain 
frequency interval which  we are  sure  contains the 
first harmonic. Then,
 we search in detail for ten 
harmonics in the frequency bands
 centered respectively at $m_{k(1)}k(i)/k(1)$
and which  amplitudes are proportional  to $\sigma_{k(1)}k(i)/k(1)$.
In each band,  we set a threshold depending on the mean and
the variance  of the power spectrum at the ends of those intervals.
We determine  the points which are over  the threshold, and from them, we
try  to determine  the nearest points to the beginning and
end of the corresponding line. One has to be careful, since  violin modes
(i.e., the transverse modes of the suspension wires)
can be present  and   stand out of
the threshold, and also, the harmonics considered can be partially under it.

 Our criteria  consists
in lowering the threshold until  there is a certain minimum number of points
over it. This number is set depending on the length of the data
(number of blocks used) and the harmonic considered. Then, we find the location
of, not the first but, a certain $n$-point, $p(n)$, that  stands out of
that threshold and  we consider  the first point of the line to be
$p(n)-n$. A similar procedure is applied for  the last point.
In case that there is just one signal present in this interval, we would
expect $p(n)-n$ to be similar to $p(1)$, but  in case that, for example, 
a violin mode is very close to the harmonic (but not overlapping with it), 
then
we hope  that  the quantity $p(n)-n$ would lie between the two lines.
Finally, the initial and final points are shifted until they reach  a local
minimum of an averaged power spectrum.

Solved these two questions, we execute our program using MATLAB  
software running
on an SGI--Origin 2000 and, after approximately 140 minutes CPU time, the
 data is stored almost clean of electrical interference.

To show  the result of  the program, in figure 4, we compare a zoom 
of the spectrogram for the frequency range between 740 and 760 Hz. 
There, we can see the  performance  of the coherent line removal algorithm,  
comparing how a line due to an harmonic of the electrical interference
 in the initial data  is removed. 
 We  notice  that the algorithm did not work properly
 for two 
sets of blocks: 1730-1793 and 3213-3276. In both cases,
this is caused 
by the 
presence of a huge number of \lq\lq bad" blocks corresponding to 
periods in which the detector is out of lock.

\begin{figure} 
\centerline{\vbox{ 
\psfig{figure=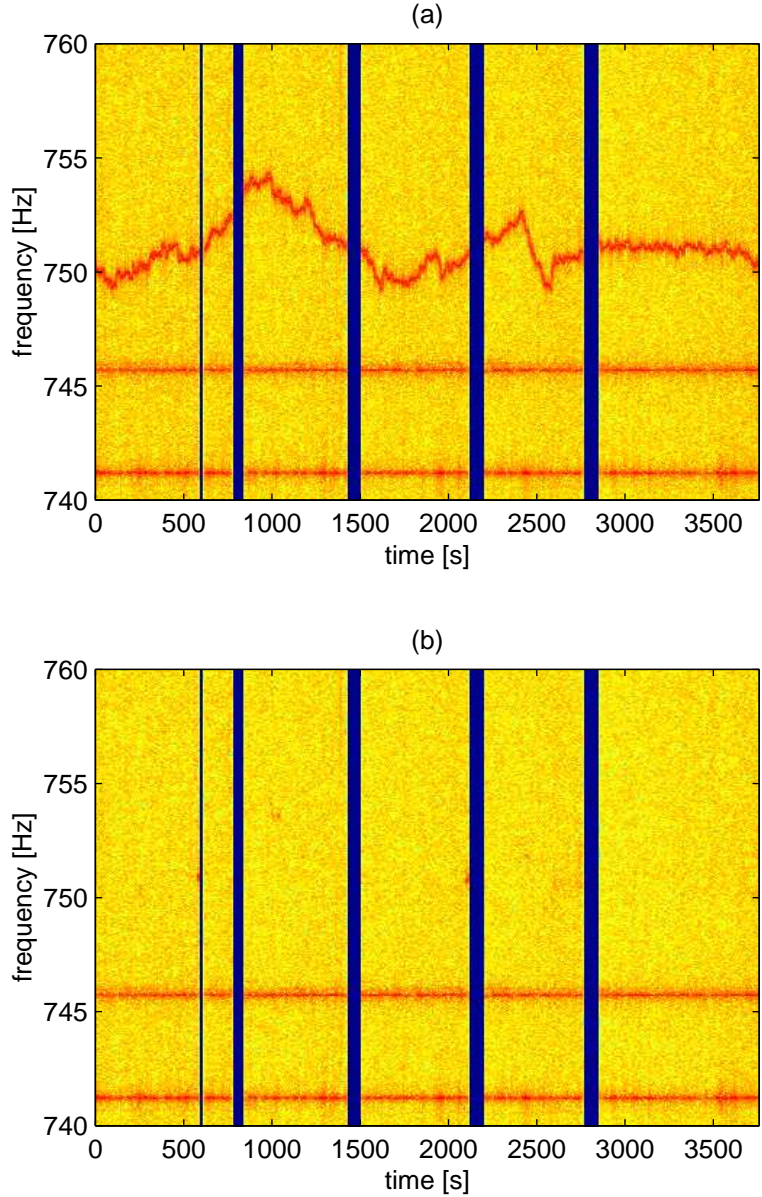,width=4.0in} 
}} 
\vspace*{10pt}
\caption[]{ Comparison of a zoom of the spectrogram.
The dark areas correspond to the periods in which the detector is
out of lock. (a) is obtained from the Glasgow data. We can observe
the wandering of the incoming electrical 
signal. The other two  remaining lines
at constant frequency correspond to violin modes. (b) The same spectrogram
as in (a) after applying coherent line removal, showing how the 
electrical interference is 
 removed to below the noise level.
} 
\vspace*{10pt}
\end{figure} 

In order to reduce the noise variance and see the quality of the performance
of the coherent line removal, we calculate 
an average periodogram using Welch's overlap method \cite{w1,w2}.
We remove the two sets of bad data,  we divide the rest 
   into sets of 128 blocks with overlaps of 64 blocks, and 
we average over the short periodograms. By doing so,
we observe how the harmonics 
 of the electricity supply frequency
 still remain undetectable over the noise level. Also,
  we discover a series of thin features
 at 701, 1002 and 1503 Hz,
 and  the presence of  violin modes at 853, 1451 and 1750 Hz
 that were buried with the electrical interference before.

\subsection{Effects on the statistics}
After removing the electrical interference, we are interested in studying 
possible side effects on the statistics of the noise in
the time domain.

For both sets of data, we calculate the mean, the standard deviation, the
skewness and the kurtosis of each block, and we compare the values obtained.
We observe that  the mean is hardly changed and its value is almost constant
around $-0.018\pm 0.002$ Volts for the \lq\lq good" blocks. 
By contrast, a big
difference  is obtained for the standard deviation. In the Glasgow data, we 
see that the standard deviation is not uniform, it tends to increase, 
specially during an on lock period  until the detector  goes out  of lock
and its average value  is around $1.50$ Volts. After the line removal, 
we observe the same behavior, but for each block, the standard deviation
 is reduced by $0.45\pm 0.05$ Volts, obtaining an average value around $1.05$
Volts (see figure 5). 

\begin{figure} 
\centerline{\vbox{ 
\psfig{figure=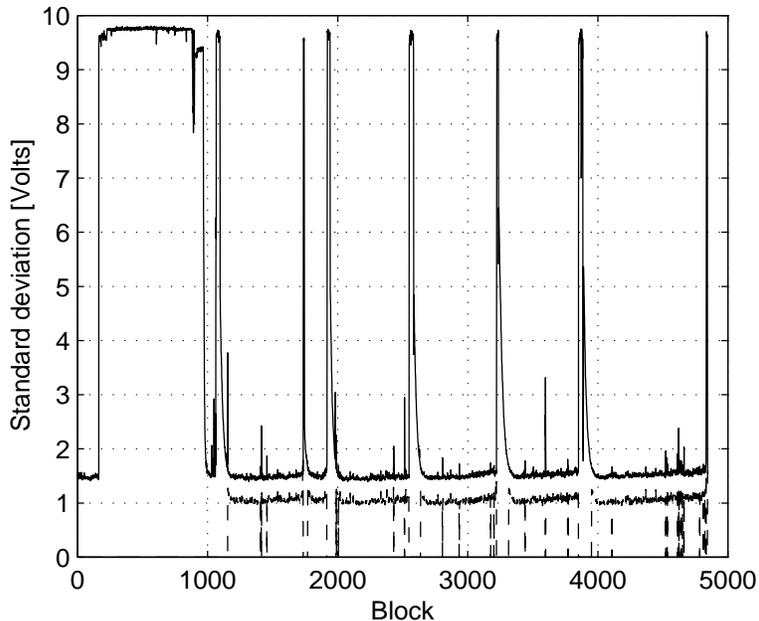,width=4.0in} 
}} 
\vspace*{10pt}
\caption[]{
The solid line (top one) corresponds to the block standard deviation
of the Glasgow data. When the detector in on lock, its typical value is
$1.50$ Volts. However, when the detector is out of lock, this value 
grows up to 10.
 The  dashed line (bottom one)  corresponds to the standard deviation
of the   data
after removing the electrical interference. Its typical value is now around 
$1.05$ Volts.
}
\vspace*{10pt} 
\end{figure} 
 The skewness characterizes the degree of asymmetry of the distribution
  around its mean. The value for each block changes  from $-0.015\pm 0.060$
  in the original Glasgow data, to $-0.005\pm 0.040$ after the line removal.
The kurtosis measures the relative peakedness or flatness of a 
distribution with respect to a normal distribution.
  In both sets of data, the
 kurtosis fluctuates from a slight negative value to a large positive
one. For the Glasgow data, the kurtosis is concentrated around
 $-0.35 \pm 0.15$, obtaining  a maximum (isolated) value  of $7.8$,
and after removing the electrical interference,
 it is around $-0.15\pm 0.15$. Therefore,
it is much   closer to the zero value, 
even though the (isolated) spikes are larger,
the highest value being $16.7$.
The kurtosis value gives an idea how Gaussian  the noise is and 
whether it has a large tail or not. A value near zero suggests a 
Gaussian nature, and a positive value indicates that the distribution
is quite peaked.
Since both values, the skewness and the kurtosis, are getting closer to 
zero after the line removal,
 we are interested in studying the possible 
 reduction of the level of non-Gaussian noise. To this end,  we choose
a set of $410$ blocks of \lq\lq good" data and we study their histogram,
calculating the number of events  that lie between different equal intervals,
 before and after removing  the
electrical interference.

Instead of plotting the number of events versus its location, we plot
their logarithm as a function of $(x-\mu)^2$, where $x$ is the
central position of the interval considered and $\mu$ is the mean
(see figure 6).
In  case that the noise distribution resembles a Gaussian, all points
should fit on a straight line of slope $-1/2\sigma^2$, where $\sigma$
is the standard deviation. We observe that this is not the case.
Although, both distributions seem to have a linear regime, they 
present a break and then a very heavy tail. The two distributions
are very different. This is mainly due to the  change of the standard 
deviation, that has a value of $\sigma_G=1.5151$ Volts for the
original Glasgow data and $\sigma_c=1.0449$ Volts after removing the
interference.

We can zoom the \lq linear' regime and change the scale in 
the abscissa  to $(x-\mu)^2/(2\sigma^2)$. Then, any Gaussian
distribution should fit into a straight line of slope -1. 
We observe that after removing the interference, it follows a 
Gaussian distribution quite well up to $4\sigma$. Whereas,
the original Glasgow data does not fit  to a straight line:
The slope changes; it does not correspond to the right one; and
close to the origin, it is flatter than a Gaussian due to the
negative kurtosis value.

\begin{figure} 
\centerline{\vbox{ 
\psfig{figure=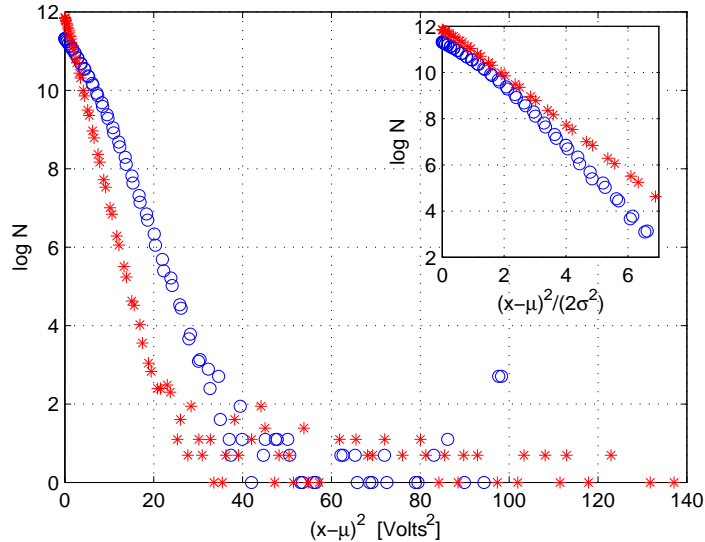,width=4.0in} 
}}
\vspace*{10pt} 
\caption[]{ Comparison of the logarithm  plot of the histogram 
for the blocks 4111 to 4520
as a function of 
$(x-\mu)^2$. The circles correspond to the Glasgow data
and the stars to the same data after removing the electrical interference.
 The Glasgow data is 
characterized by $\mu=-0.0182$ Volts and $\sigma = 1.5151$ Volts.
After the line removal, we obtain the 
values of  $\mu=-0.0182$ Volts and $\sigma = 1.0449$ Volts.
In the right-hand corner, there is zoom  of the original figure, but
rescaled so that
the abscissa corresponds to $(x-\mu)^2/2\sigma^2$.
If the data  resembles a Gaussian distribution, 
we will expect a single straight line of slope -1. This is not the
case for the Glasgow data, but it seems to be satisfied  for the
clean data up to 
$4\sigma$. 
The large number of points in the highest bin of the Glasgow data is
an effect of saturating the ADC. These points are spread to higher and
lower 
voltages by line removal.
}
\vspace*{10pt} 
\end{figure} 

We can study the population of the noise in the tail of the distributions
and see how it is affected by the line removal. The number of events
that exceeds the $\pm 7$ Volts is of 60 in the Glasgow data and of
51 after removing the interference, meaning that the events are now
concentrated at lower voltages. But, one can be more interested
in comparing the number of events that exceeds 
a certain standard deviation  value.  The population outstanding
 $5\sigma$ is just 51 for the Glasgow data, but of 96 after the line
 removal. This higher number of events in the tail is due to the big
 reduction  of the  standard deviation value, since the same 
 $\sigma$ value corresponds to a much lower voltage.
 
The Glasgow data has a dynamical range from $-10$ to $+10$ Volts, 
as it was defined by the computer board they use.
After removing the electrical interference, we observe  events at
11.8 Volts. This implies that the amplitude of the
signal we remove is of at least 1.8 Volts.
This high signal amplitude, together with the previous 
existence of events close to the limit of the dynamical range allowed,
and the decrease  of the standard deviation can explain how,
in some \lq isolated' cases, the kurtosis value grows so much.
%%%%%%%%%%%%%%%%%%%%%%%%%%%%%%%%%%%%%%%%%%%%%%%%%

In order  study the Gaussian character, we apply two statistical tests to the
data: the chi-square test \cite{chi} that measures  the discrepancies
between binned distributions,  and the one-dimensional Kolmogorov-Smirnov
test \cite{ks} that measures the differences between  cumulative 
distributions of a continuous data. Both tests can be applied  to our data
 since
we can always turn continuous data  into binned data, 
by grouping the events into specified ranges of the continuous variable.

We compute the significance  probability for each block of the data
using both tests and we check whether the distribution are Gaussian or not.
The two tests are not equivalent but in any case, the values of the
significance probability would be close to unity for 
distribution resembling a Gaussian distribution (see figures 7 and 8).

\begin{figure} 
\centerline{\vbox{ 
\psfig{figure=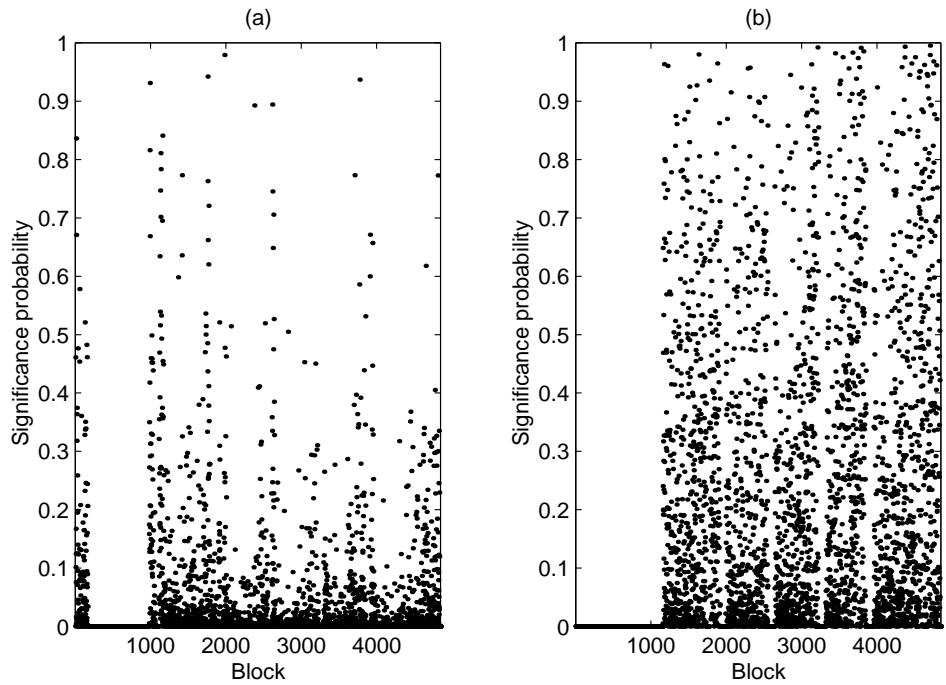,width=5.0in} 
}} 
\vspace*{10pt}
\caption[]{ Comparison of the block
 chi-square test: (a) for the Glasgow data, (b)  
after the removal of the interference.
} 
\vspace*{10pt}
\end{figure} 
\begin{figure} 
\centerline{\vbox{ 
\psfig{figure=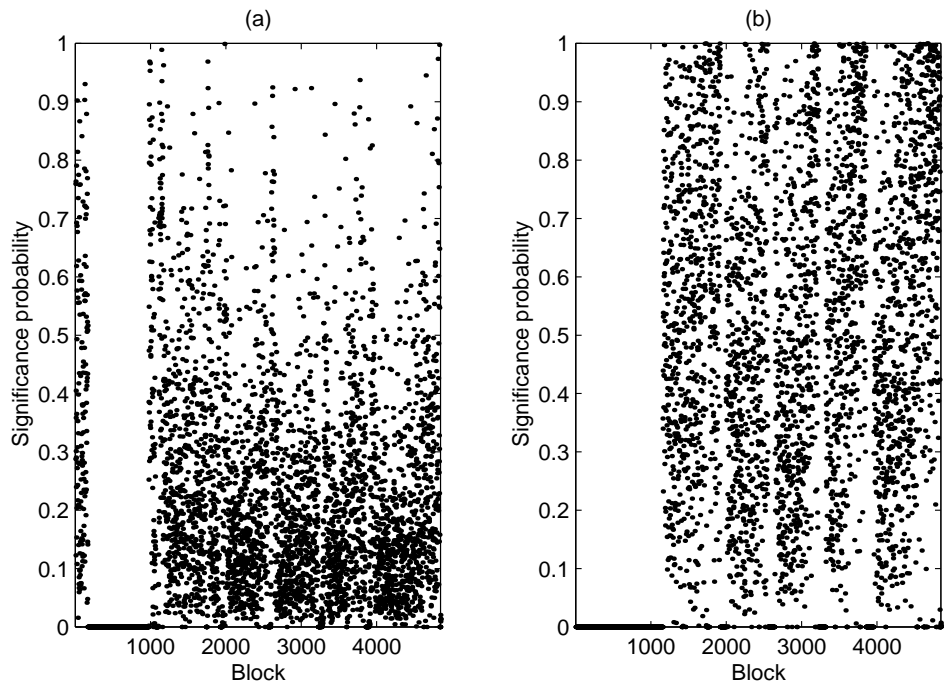,width=5.0in} 
}} 
\vspace*{10pt}
\caption[]{Block Kolmogorov-Smirnov test: (a) for the Glasgow data,
(b) 
after the removal of the interference.
} 
\vspace*{10pt}
\end{figure} 

In all cases, we find that the data deviates from a Gaussian distribution 
by a wide margin. Although, for some blocks we get a high significance 
probability, this cannot be taken as the dominant trend. 
As a result of both tests, we note how the significance  probability increases
after removing the electrical interference, showing  that this
procedure suppresses some non-Gaussian noise, although, generally 
speaking, the distribution may still be described  as being 
non-Gaussian in character.

\section{Discussion}
We have described a new exciting technique for removing troublesome
interference from external coherent signals (such as the electricity
supply frequency and its harmonics) which can obscure wide frequency
bands.  This approach can leave the spectrum clean enough to see
true gravitational waves that have been buried in the  interference.
Therefore, this new method  appears to be good news  as far as searching 
for continuous waves is concerned.

We have applied   {\it coherent line removal}  to the Glasgow 
laser interferometer data and we have succeeded in removing the 
electrical interference. By doing so,  we have discovered some new thin 
features at 701, 1002 and 1503 Hz, and also some violin modes at 
853, 1451 and 1750 Hz that were masked by the interference in the 
original data.

The results of the previous section  indicate that {\it coherent line removal}
 can also reduce 
the level
of non-Gaussian noise. This result
 is very encouraging for large-scale
interferometers.
Nicholson et al. \cite{n} reported that the effective strain sensitivity in 
coincident observations for short bursts in the time scale, was a factor of 2
worse
than the theoretical best limit that the detectors could have set, the
excess being due to unmodeled non-Gaussian noise, and they also indicate
that reducing this non-Gaussian noise
 could raise the sensitivity and duty
cycle of working  detectors close to their optimum performance.

{\it Coherent line removal} can be applied to any kind  of 
coherent interference signal. It only requires  phase-coherence between the 
fundamental and several harmonics.
Since the  algorithm can be applied recursively for small pieces  of
a long data stream, 
the global physical process that produces  the interference (and all
the harmonics) does not need to be stationary, i.e., the parameters 
$a_n$ in Eq.~(\ref{e3}) may change  in time. We only  require that
the process could be  considered stationary  for those time scales
in which the algorithm is applied. For every small piece of data, the
reference function $M(t)$  and all the parameters $\rho_n$ are 
calculated allowing any  possible changes.
This method can be considered as  
an adaptive method that has
the ability  to adjust its own parameters automatically and requires
little knowledge of the signal we want to remove.

{\it Coherent line removal}
 may have more applications, not only for the detection of 
 {\sc gw} radiation, but also, for example, in radioastronomy
 or in other completely different fields.

We hope that the results described before will encourage others
in testing and applying this new technique to other problems.

%\section*{Acknowledgments}
\acknowledgements{
We would like to thank C. Cutler, A. Kr\'olak, M.A. Papa and
A. Vecchio for helpful discussions,
and  J. Hough and the gravitational waves group at Glasgow University
for providing their gravitational wave interferometer data for analysis.
This work was partially supported by the European Union, 
 TMR Contract
No. ERBFMBICT972771.}

 %\end{thebibliography}

\end{document}